\documentclass[journal]{IEEEtran}

\usepackage{cite}
\usepackage{siunitx}
\usepackage{hyperref}
\usepackage{todonotes}
\usepackage{acronym}
\usepackage{multirow}

\newacro{ISA}[ISA]{International Society of Automation}
\newacro{TSCH}[TSCH]{time-slotted channel hopping}
\newacro{MAC}[MAC]{medium access control}
\newacro{6TiSCH}[6TiSCH]{IPv6 over the TSCH mode of IEEE 802.15.4e}
\newacro{RPL}[RPL]{routing protocol for low-power and lossy networks}
\newacro{IWSN}[IWSN]{industrial wireless sensor network}
\newacro{JRQ}[JRQ]{join request}
\newacro{JRS}[JRS]{join response}
\newacro{WuR}[WuR]{Wake-up Radio}
\newacro{KA}[KA]{keep-alive}
\newacro{IoT}[IoT]{Internet of Things}
\newacro{WPT}[WPT]{wireless power transfer}
\newacro{RF}[RF]{Radio Frequency}
\newacro{CCA}[CCA]{Clear Channel Assessment}

\newacro{}[]{}
\begin{document}

\title{Integrating Battery-Less Energy Harvesting Devices in Multi-hop Industrial Wireless Sensor Networks}

\author{Dries~Van~Leemput, Jeroen~Hoebeke, Eli~De~Poorter
\thanks{D. Van Leemput, J. Hoebeke, and E. De Poorter are with IDLab, Department of Information Technology, Ghent University - imec, Belgium e-mail: firstname.lastname@ugent.be.}
\thanks{© 2024 IEEE.  Personal use of this material is permitted.  Permission from IEEE must be obtained for all other uses, in any current or future media, including reprinting/republishing this material for advertising or promotional purposes, creating new collective works, for resale or redistribution to servers or lists, or reuse of any copyrighted component of this work in other works.}}

\markboth{IEEE Communications Magazine}%
{Van Leemput \MakeLowercase{\textit{et al.}}: Strategies for including energy harvesting devices in multihop IIoT wireless networks}

\IEEEaftertitletext{\vspace{-1.1\baselineskip}}

\maketitle

\begin{abstract}
Industrial wireless sensor networks enable real-time data collection, analysis, and control by interconnecting diverse industrial devices. In these industrial settings, power outlets are not always available, and reliance on battery power can be impractical due to the need for frequent battery replacement or stringent safety regulations. Battery-less energy harvesters present a suitable alternative for powering these devices. However, these energy harvesters, equipped with supercapacitors instead of batteries, suffer from intermittent on-off behavior due to their limited energy storage capacity. As a result, they struggle with extended or frequent energy-consuming phases of multi-hop network formation, such as network joining and synchronization. To address these challenges, our work proposes three strategies for integrating battery-less energy harvesting devices into industrial multi-hop wireless sensor networks. In contrast to other works, our work prioritizes the mitigation of intermittency-related issues, rather than focusing solely on average energy consumption, as is typically the case with battery-powered devices. For each of the proposed strategies, we provide an in-depth discussion of their suitability based on several critical factors, including the type of energy source, storage capacity, device mobility, latency, and reliability.
\end{abstract}


\IEEEpeerreviewmaketitle

\section{Introduction}
\IEEEPARstart{T}{he} use of \acp{IWSN} has shown to be a viable alternative to wired solutions for industrial applications due to their scalability, cost-effectiveness, and the possibility to be deployed in hard-to-reach areas within industrial sites. However, devices in \acp{IWSN} are often mains-powered, hindering easy roll-out and cost-effective network expansions, or battery-powered, which requires regular battery replacements as the batteries last often less than a few years. In addition, replacing batteries in hard-to-reach areas is often dangerous, costly, or impossible. Moreover, batteries are hazardous, preventing them from being deployed in strictly regulated areas such as potentially explosive atmospheres. This requires ATEX-certified batteries, which significantly increase the deployment and maintenance cost. As a result, there is a need to power devices in regulated and hard-to-reach areas that do not require regular and dangerous maintenance. There has been a significant interest in powering \ac{IoT} devices using energy harvesters in combination with supercapacitors, which experience better temperature performance, have a significantly longer lifetime than batteries, are more resistant to current peaks, and do not pose the risk of explosion when overcharged or short-circuited. In addition, it removes the need for maintenance as these battery-less devices are deploy-and-forget. As a result, integrating battery-less devices in \acp{IWSN} may be the solution to provide complete and safe coverage of the entire industrial site. Depending on the deployment environment, the energy harvesting source can be solar, kinetic, vibration, thermal, \ac{WPT}, etc.

However, integrating battery-less devices in \acp{IWSN} does come with its challenges. The available harvesting power is often variable or extremely limited and while supercapacitors have significant benefits over batteries, they also have a lower energy density and higher self-discharge rate. This results in an intermittently on-off behavior of the battery-less device. In contrast, existing \ac{IWSN} standards such as WirelessHART, ISA100.11a, 6TiSCH, etc. all assume the presence of a stable power supply. While existing scientific literature focuses on reducing the average energy consumption of \acp{IWSN}, this intermittent behavior is typically not supported. Nonetheless, this behavior is especially challenging for two network phases: the joining procedure and the synchronization procedure, both of which are strongly hindered by intermittent on-off behavior. This article bridges this gap by presenting three strategies to integrate battery-less devices in \acp{IWSN} while taking the intermittent nature into account. Specifically, the main contributions of this article are:
\begin{itemize}
    \item We identify the main challenges for integrating battery-less devices into IWSNs, emphasizing the intermittent behavior due to limited energy storage rather than focusing on the average energy consumption: (i) executing a energy-intensive joining phase and (ii) a frequent synchronization phase.
    \item We propose three strategies that address these challenges when integrating battery-less devices. These strategies either (i) optimize these phases to align with the intermittent behavior or (ii) remove them by transferring the workload to routers.
    \item We assess each strategy's impact on various energy and network performance metrics, including energy storage capacity, energy availability, mobility, latency, and reliability. Our analysis allows us to identify the most suitable use case scenarios for each integration strategy.
\end{itemize}

The remainder of this article is structured as follows: Section \ref{sec:multihop_IWSNS} lists and compares the current \ac{IWSN} standards and the challenges that intermittency poses for successful integration of battery-less devices. Sections \ref{sec:synchronous}, \ref{sec:ad-hoc}, and \ref{sec:asynchronous} propose three strategies, including optimizations in terms of joining and synchronization phases. Section \ref{sec:evaluation} compares the strategies in terms of energy harvesting and performance capabilities. Finally, Section \ref{sec:conclusion} concludes the paper with a conclusion and future directions.

\section{Multi-hop Industrial wireless sensor networks and Intermittency}
\label{sec:multihop_IWSNS}
The \ac{ISA} classifies industrial applications according to their maximum allowed latency and other operational requirements, ranging from control and monitoring applications (classes 4-6) with non-real-time and relaxed reliability requirements; to safety and control applications (classes 1-3) with strict latency and reliability requirements \cite{Das2017,Raza2018}. For the former application classes, contention-based wireless network protocols such as ZigBee, Bluetooth Mesh, and Matter are able to meet these relaxed requirements. Furthermore, their contention-based nature reduces the complexity of synchronization and joining procedures. However, these protocols lack the latency and reliability guarantees required for safety and control applications. Therefore, several \ac{IWSN} standards have been proposed for critical industrial applications, such as WirelessHART, ISA100.11a, and \ac{6TiSCH}. They have similar characteristics in terms of medium access, routing, synchronization, and joining procedures \cite{WirelessHART,ISA100.11a,Vilajosana2020}. To meet the strict latency and reliability requirements, these protocols employ deterministic and scheduled communications based on a\ac{TSCH} \ac{MAC} layer, which combines time and frequency diversity by using a predefined hopping sequence. Time is divided into timeslots, which can be used exclusively between two devices (dedicated slots) or shared between multiple devices (shared slots) to allow multicast frames for synchronization and network control. Through frequency hopping, these protocols are able to operate in harsh industrial environments with a high degree of frequency-selective fading. Depending on the standard, the hopping sequence is either static or dynamically adapted to the network requirements.

In complex, large-scale industry environments, deploying devices over multiple hops is often necessary to achieve the desired site coverage. As a result, WirelessHART \cite{WirelessHART}, ISA100.11a \cite{ISA100.11a}, and 6TiSCH \cite{Vilajosana2020} all provide meshing support. While the routing and time scheduling in WirelessHART and ISA100.11a are centrally managed by a network manager\cite{WirelessHART,ISA100.11a}, \ac{6TiSCH} makes use of a hierarchical routing protocol: the \ac{RPL}. All three protocols differentiate between border routers (that connect the \ac{IWSN} to a network backbone or other networks), routers (able to forward traffic), and end devices (which form the leaves of the network tree). All standards require joining nodes to perform multi-hop authentication with the border router by a \ac{JRQ} and \ac{JRS}. Naturally, end devices consume less energy than their router counterparts, which makes them an interesting option to power them with energy harvesters. \textbf{As industrial networks often combine both critical (class 1-3) and monitoring (class 4-6) applications, we aim to allow the integration of intermittent battery-less devices into existing class 1-3 \acp{IWSN}.} Nonetheless, several challenges need to be addressed.

\begin{figure}
    \centering
    \includegraphics[width=0.9\linewidth]{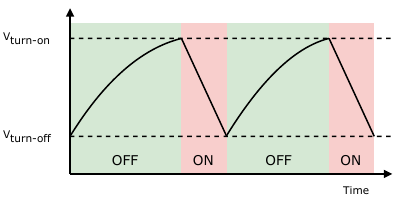}
    \caption{Intermittent on-off behavior of a battery-less device. The supercapacitor is charged by the energy source to the turn-on threshold, permitting the device to turn on and execute tasks. This depletes the supercapacitor until the turn-off threshold is reached, forcing the device to power off and allowing the supercapacitor to be recharged by the energy source.}
    \label{fig:intermittency}
\end{figure}

While \ac{TSCH}-based network protocols meet the strict communication and operational requirements of the lower \ac{ISA} application classes, their complex bootstrap and network-management procedures make integrating battery-less devices a challenging task \cite{Das2017,Chew2021}. After all, battery-less devices generally make use of a supercapacitor as the storage element, resulting in an intermittent on-off behavior of the device due to the lower energy density compared to batteries. Fig. \ref{fig:intermittency} depicts this intermittent on-off behavior, where a supercapacitor is charged to $V_{turn-on}$, allowing the device to perform some tasks and depleting the capacitor to $V_{turn-off}$. This forces the device to turn off as there is not sufficient energy stored to power the device. Afterward, an energy source can recharge the capacitor. Depending on the device, application, and available harvesting power, the device can be forced to switch off when reaching $V_{turn-off}$ or deliberately switch off / enter a sleep state before reaching this threshold. Although the latter option appears to be favorable and can be achieved by intelligent task-scheduling algorithms, the power variability of energy harvesters either makes this option obsolete or at least not guaranteed. In any case, \textbf{the capacitor should be able to hold enough energy to execute the highest energy consumption period. As such, the duration and frequency of the most energy-intensive and frequently occurring periods pose the highest restrictions on intermittent devices, rather than the average energy consumption as is the case for battery-powered devices} \cite{VanLeemput2023}.

To accommodate this requirement, end devices can tune the frequency with which they report data depending on the available harvesting power, taking into account the intermittent behavior. However, application developers cannot change the network behavior. As such, \textbf{for \acp{IWSN}, the intermittent behavior of battery-less devices presents two main challenges for their integration in \acp{IWSN}: devices need to execute an energy-intensive} joining procedure and have to periodically synchronize to nearby routers. Due to the similarities of these phases across \ac{IWSN} standards, the challenges for integrating battery-less devices apply to WirelessHART, ISA100.11a, and 6TiSCH, opening up opportunities to design general strategies suitable for most relevant industry standards. In the next sections, we will provide three such strategies that address the aforementioned challenges for integrating battery-less devices into \acp{IWSN}. An overview and comparison of the integration strategies can be found in Table \ref{tab:comparison}, listing which challenging procedures are included and optimized, typical energy requirements, and the performance impact on both battery-less devices and routers. 

\setlength{\textfloatsep}{10pt}

\begin{table*}[]
    \centering
    \def\arraystretch{1.2}
    \caption{Comparison of integration strategies indicating: which phases they include, possible optimizations of those phases, typical energy requirements, and performance impact on battery-less devices and routers.}
    \begin{tabular}{l l | l l l}
          &&  \textbf{Strategy 1: Synchronized} & \textbf{Strategy 2:} & \textbf{Strategy 3: Non-synchronized} \\
          &&  \textbf{communication} & \textbf{Ad-hoc joining} & \textbf{communication} \\
         \hline
         \multicolumn{2}{c|}{Joining phase} & yes & yes & no \\
         \multicolumn{2}{c|}{Synchronization phase} & yes & only once & no \\
         \hline
         \multicolumn{2}{c|}{\multirow{4}{*}{Potential optimizations}} & Limited advertisement channels, & Wake-up radios, beacon & \\
         && supercapacitor sizing, & solicitation, duty-cycled joining, & Multi-interface routers, \\
         && scheduled joining, keep alive frames, & hierarchical management, & supercapacitor sizing \\
         && adaptive synchronization & supercapacitor sizing & \\
         \hline
         \multicolumn{2}{c|}{Standard compliance} & yes & yes & no \\
         \hline
         & Energy source & Consistent $>\SI{250}{\micro\watt}$ & Variable $> \SI{150}{\micro\watt}$ & Variable $>\SI{50}{\micro\watt}$ \\
         Battery-less device & Storage capacity & Sufficient to join $>\SI{100}{\milli\farad}$ & Sufficient to join $>\SI{100}{\milli\farad}$ & Sufficient to transmit $>\SI{100}{\micro\farad}$ \\
         characteristics & Mobility & Limited & Mobile & Mobile \\
         & Performance & Low latency $<T_{sf}$& High latency $>T_{join}$& Low latency $<T_{slot}$ \\
         \hline
         Router & Energy source & Battery-powered & Mains-powered & Mains-powered \\
         characteristics & Performance & Unaffected & Reduced performance peaks & Reduced overall performance \\
         \hline
    \end{tabular}
    \label{tab:comparison}
\end{table*}

\section{Strategy 1: synchronized communication}
\label{sec:synchronous}
\begin{figure}
    \centering
    \includegraphics[width=\linewidth]{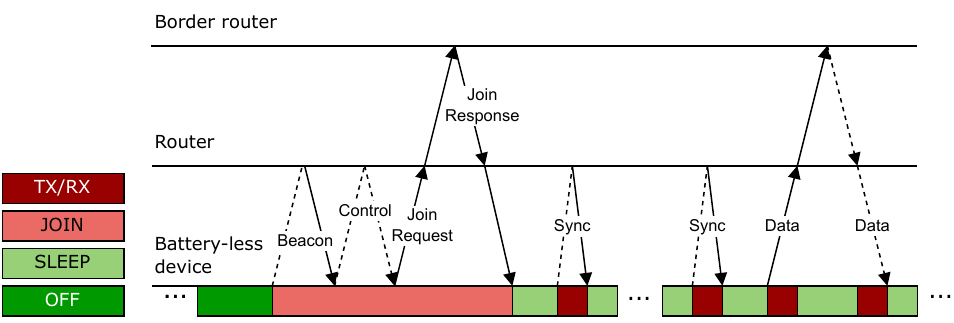}
    \caption{\textbf{Synchronized communication including joining and synchronization}. Battery-less devices join the network during bootstrap and periodically synchronize with a nearby router. Joining comprises scanning for beacons, protocol-dependent control messages, and multi-hop authentication with the border router (including a Join Request and Join Response required in all standards). Data transmission and optional reception occur during dedicated timeslots.}
    \label{fig:synchronous}
\end{figure}

The most straightforward way to incorporate battery-less devices in \acp{IWSN} is to retain the joining and synchronization phases. That way, minimal to no changes are required to battery-less devices and routers. Battery-less nodes join the network during bootstrap as end devices and participate in the \ac{TSCH} schedule, periodically synchronizing to nearby routers. As such, the existing schedule can be used to transmit and/or receive data, benefiting from the protocol's performance in terms of latency and reliability. This approach is depicted in Fig. \ref{fig:synchronous}, which shows the interaction between a battery-less node, a router, and the border router, where dotted lines represent optional messages. We distinguish between four different power states of the end device: TX/RX, join, sleep, and off. The joining state comprises scanning or soliciting for beacons, protocol-dependent message exchanges with nearby routers, and multi-hop authentication with the border router. Once fully joined, the battery-less node periodically synchronizes to a nearby router and uses dedicated TX and RX timeslots for reliable and latency-bound communication.

\setlength{\textfloatsep}{10pt}

\subsection{Joining Optimizations}
As shown in Table \ref{tab:comparison}, this approach retains the joining and synchronization phases. Therefore, it is important to reduce their energy consumption to the fullest extent. Joining a \ac{TSCH} network requires an energy-intensive scanning phase where devices scan for beacons, which advertise the network and contain the deployed schedule. Once one or several beacons have been received, the device is joined to the \ac{TSCH} schedule. However, due to channel-hopping, receiving a beacon can take a significant amount of time, as the device must be listening on the same channel as a neighboring node is transmitting the beacon on. Assuming a single advertisement slot in a superframe with a duration $L_{sf}$, a device may need to wait for $N_{ch}*L_{sf}$ before receiving a beacon on the channel it is listening, with $N_{ch}$ representing the number of channels. After joining the \ac{TSCH} schedule, multiple control messages need to be exchanged with local neighbors and a network manager to join the network fully, establish routes, and complete authentication \cite{WirelessHART,ISA100.11a,Vilajosana2020}. This increases the energy consumption and duration of the joining procedure, especially if the network manager is located multiple hops away, which reduces the probability of a battery-less device successfully (re)joining the network. If the network manager is located four hops away, completely joining a network can last 80 seconds, assuming a one-second superframe \cite{Das2017}. In comparison, charging a \SI{0.33}{\farad} supercapacitor with vibration energy can last twelve hours, which is only sufficient to power a device for twelve seconds during a normal joining procedure \cite{Chew2021}. 

As such, optimizing the always-on scanning phase is crucial for a successful joining attempt. In fact, devices consume approximately twice more energy during scanning compared to negotiation with the border router and/or network manager \cite{Chew2021}. To ensure sufficient capacity to join the network, the following optimizations can be considered, listed in Table \ref{tab:comparison}. (i)
\textbf{Limiting the number of advertisement channels} can significantly reduce scanning time \cite{Chew2021,Das2017,Kalita2022}. While this negatively impacts resilience against frequency-selective fading, more channels can be used for data and other control traffic. It is also crucial to maximize the probability of successfully joining the network in one attempt, as it might require a significant period to recharge a supercapacitor again for a second attempt. Therefore, (ii) appropriately \textbf{sizing the supercapacitor} \cite{Chew2021} can increase the joining probability. Assuming a \SI{100}{\milli\farad} supercapacitor, the recharge time is kept reasonable for a harvesting power as low as \SI{100}{\micro\watt} \cite{Chew2021}. In addition, environmental information can be used for (iii) \textbf{scheduling joining} during periods of high power availability (e.g., during a bright interval when using solar harvesting), which also increases a successful joining probability.

\subsection{Synchronization Optimizations}
Even if devices are able to join, they need to remain synchronized with the schedule by receiving beacons and maintain the network by exchanging control frames with neighbors. While this is not as energy-intensive as joining the network (a typical energy consumption peak in stable operation is around \SI{1.5}{\milli\joule} \cite{VanLeemput2023} compared to \SI{325}{\milli\joule} to fully join a \ac{TSCH} network \cite{Das2017}), the requirement of synchronizing frequently is not evident for intermittent devices. As such, after some time battery-less devices might disconnect from the network due to energy depletion, forcing them to perform the energy-intensive joining procedure again once sufficient energy has been harvested. This not only consumes a significant amount of energy from the battery-less device but also floods the network with control traffic, leading to the increased energy consumption of neighboring devices and reduced network performance. Therefore, maximizing the synchronization interval and limiting the energy consumption of a synchronization task can increase integration opportunities. The synchronization interval is dictated by clock accuracy and the guard time to account for the worst-case clock drift during message exchanges. A higher clock accuracy results in an increased synchronization interval at the cost of higher-priced clock crystals. On the other hand, increasing the guard time also increases the synchronization interval but inevitably increases the energy consumption of a single message exchange as devices need to listen longer for incoming frames \cite{Chang2015, Vilajosana2014}.

To increase the synchronization interval, (i) an \textbf{adaptive synchronization}, can be employed, which measures and models the relative clock drift between devices instead of considering the worst-case clock drift \cite{Chang2015}. Independent of the synchronization interval, \ac{TSCH} provides two methods for synchronizing devices: using beacons received from neighboring routers or sending \ac{KA} frames periodically and using timing information from the associated acknowledgment. For energy harvesting devices, (ii) \textbf{using \acp{KA}} is more efficient since devices are not required to listen for incoming beacons but transmit an empty \ac{KA} themselves whenever needed. Next to periodical synchronization, \acp{IWSN} require other control traffic to manage and maintain the network. Luckily, WirelessHART, ISA100.11a, and \ac{6TiSCH} all provide the option of joining the network as an end device only, indicating it is not able to perform routing duties. Hence, the control traffic of these devices can be kept to a minimum without altering the communication protocol. As a result, the best option to integrate energy harvesting devices is to define them as end devices.

\subsection{Use cases}
An important benefit of this approach is that it poses minimal restrictions on the rest of the network, as the existing schedule can be used. In addition, this approach is compliant with existing \ac{IWSN} standards, as shown in Table \ref{tab:comparison}. This ensures the network performance is unaffected by integrating battery-less devices and enables routers to be battery-powered. In terms of the battery-less device, only one energy-intensive joining procedure must be executed during bootstrap as the device stays synchronized. As a result, Table \ref{tab:comparison} indicates that this approach requires a relatively consistent energy generated by the energy harvester as the node needs to synchronize periodically. A continuous harvesting power to support a stable network operation should not drop below \SI{250}{\micro\watt}.

In addition, the storage element needs to be sized sufficiently large to cover periods of reduced power generation and to deliver a sufficient energy peak during the initial joining procedure. A minimal capacitance of \SI{100}{\milli\farad} was found to be appropriate to successfully join a \ac{TSCH} network in the first attempt while using an optimized joining procedure \cite{Chew2021}. Furthermore, especially when using active synchronization, mobility is limited unless synchronizing with multiple routers, which inevitably increases energy consumption. On the other hand, assuming a relatively stable harvesting power, an appropriately sized supercapacitor, and static battery-less devices, this approach can enable strict communication requirements in terms of latency and reliability because the node is participating in the \ac{TSCH} schedule, benefiting from its characteristics. However, latency is bounded by the slotframe duration $T_{sf}$, which can be decreased to improve latency if the harvesting power reaches a sufficient level. On the other hand, $T_{sf}$ also depends on the required synchronization interval, determined by the clock accuracy of the available hardware platform.

\section{Strategy 2: Ad-hoc joining}
\label{sec:ad-hoc}
\begin{figure}
    \centering
    \includegraphics[width=0.9\linewidth]{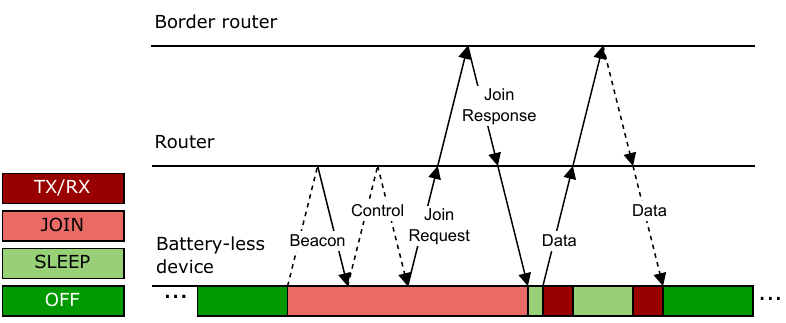}
    \caption{\textbf{Ad-hoc joining without periodical synchronization}. Whenever the battery-less device needs to send an update or sufficient energy is available, it rejoins the network until data is transmitted/received. Afterward, it disconnects and powers off to save and harvest energy.}
    \label{fig:ad-hoc_joining}
\end{figure}

If the above requirements for a battery-less node to synchronize periodically cannot be guaranteed, an ad-hoc joining approach may be better suited to integrate the device into the network. As such, this approach is compliant with existing \ac{IWSN} standards, as shown in Table \ref{tab:comparison}. This strategy removes the need for periodical synchronization but still requires nodes to join the network and synchronize once. The principle is depicted in Fig. \ref{fig:ad-hoc_joining}, using the same power states as Fig. \ref{fig:synchronous}. For the majority of the time, the device can be powered off while harvesting energy from its environment. Whenever sufficient energy is available, the device can rejoin the network and provide and/or receive an update using temporary dedicated TX and/or RX slots with a nearby router. 

\setlength{\textfloatsep}{10pt}

\subsection{Joining Optimizations}
Since the battery-less node will join the network each time it needs to communicate, Table \ref{tab:comparison} lists more drastic methods to reduce the energy consumption of the joining process. (i) Using a \textbf{\ac{WuR}} to wake up a battery-less node while simultaneously transmitting a beacon could be very effective as the device does not need to passively listen for incoming beacons. This, however, requires the device and router to agree on the advertising channel in advance. (ii) Another energy-efficient method to join a \ac{TSCH} network is to join actively by transmitting a \textbf{beacon solicitation}, which triggers the transmission of a beacon by neighboring routers. Such an approach is available in ISA100.11a, called active scanning, but it inevitably increases energy consumption of routers since they have to listen for beacon solicitations during idle periods \cite{ISA100.11a}. (iii) Similar to the synchronized communication strategy, appropriately sizing the storage element to successfully join the network in the first attempt significantly increases the joining probability.

While the scanning phase is the most crucial period to be optimized in the joining process, optimizing the multi-hop authentication and additional procedures can further increase the probability of successfully joining the network. (iv) Employing a \textbf{hierarchical network management} scheme aims to achieve this optimization by delegating some responsibilities of the network manager to the routers, resulting in a faster overall (re)joining time for end devices, especially benefiting devices that frequently lose connectivity due to energy depletion \cite{Das2017}. (v) Another option is to use a \textbf{duty-cycled joining process}, where devices attempt to join while other devices wait, reducing their energy consumption. This is of particular importance when multiple devices try to join the network at the same time and has been shown to successfully allow battery-less nodes to join a \ac{TSCH} network \cite{Chew2021}.

\subsection{Use cases}
The obvious benefit of this approach is the lack of synchronization, allowing the device to power off completely and harvest energy. As shown in Table \ref{tab:comparison}, this is particularly favorable for variable harvesting power patterns, as the storage element will eventually recharge irrespective of the power stability. A harvesting power of \SI{150}{\micro\watt} is sufficient to start charging a correctly sized \SI{100}{\milli\farad} supercapacitor to join the network. In addition, the device can be mobile as it can rejoin with any nearby router. However, because the joining procedure is by far the most energy-intensive period, it may take a long time to charge the storage element sufficiently to undertake a joining attempt. After all, the necessary scanning duration before receiving a beacon is not fixed, making it possible to deplete the storage element entirely without even successfully joining the network. This leads to unpredictable behavior and very low update rates, impacting latency and reliability. Even for very high harvesting power, the latency is bounded by the joining time $T_{join}$.

Moreover, each joining process reduces the performance of other network devices. Control traffic, used for authentication and updating the network state, not only floods the network during joining but also when a device leaves the network. This results in reduced performance during these periods as resources are allocated to joining devices rather than data traffic. As a result, from a router's perspective, the number of joining processes should be kept to a minimum. This requires either restricting the update rate of ad-hoc joining devices or limiting the overall number of such devices. Furthermore, beacon solicitation or \acp{WuR} draw more energy from routers, increasing their energy consumption and requiring them to be mains-powered for this approach (Table \ref{tab:comparison}).

Nonetheless, this approach is suitable for battery-less nodes with a low update rate, where synchronization is either not worth the energy, given the low update rate, or impossible because of the high harvesting power variability. Therefore, it provides an opportunity for energy harvesting devices that lack a stable energy source to connect to the network.

\section{Strategy 3: Non-synchronized communication}
\label{sec:asynchronous}
\begin{figure}
    \centering
    \includegraphics[width=0.8\linewidth]{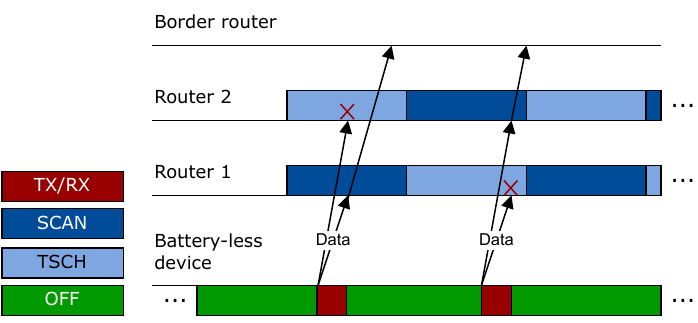}
    \caption{\textbf{Non-synchronized communication without joining and synchronization}. The EH node does not join the network, but posts data non-synchronized and powers off immediately. Nearby routers alternate between scanning and \ac{TSCH} to provide full coverage in time and frequency at each location to receive data, requiring tight coordination and dense network deployment.}
    \label{fig:asynchronous}
\end{figure}

In some scenarios, even the ad-hoc joining approach may not suffice to integrate battery-less devices in an \ac{IWSN}. Therefore, to limit energy consumption in the battery-less node, a non-synchronized communication approach removes both the joining and synchronization phases. Fig. \ref{fig:asynchronous} shows the basic principle, where the battery-less node is turned off until it has enough energy to send an update, but does not join the network. Instead, it transmits its data non-synchronized, expecting it to be received by a nearby router. This means nearby routers cannot use the existing \ac{TSCH} schedule as it is, because the battery-less node is unaware of this schedule. Alternatively, routers need to alternate between the \ac{TSCH} schedule and a scanning period, during which they can receive non-synchronized frames. However, due to the channel hopping feature, sufficient routers need to be available so every channel is covered at all times at the location of the battery-less node. Such an approach was previously implemented and validated in \cite{Das2017}, making use of the slow hopping feature of ISA100.11a. However, implementing it in WirelessHART or \ac{6TiSCH} would require modifications to the standard as the slot size and channel hopping sequence are fixed \cite{WirelessHART,Vilajosana2020}.

\setlength{\textfloatsep}{10pt}

\subsection{Implementation Challenges}
This strategy poses significant demands on the network, requires precise coordination and dense deployment. Ensuring optimal performance for a battery-less device requires more than $N_{ch}$ routers within its communication range, where $N_{ch}$ represents the number of channels used by the battery-less device. If only $N_{ch}$ routers are reachable, they must remain in a constant listening state, precluding participation in the \ac{TSCH} schedule and hindering mesh network activities. To restrict the impact to a \SI{50}{\percent} reduction in router performance, at least $2N_{ch}$ routers should be within communication range, allocating half of their duty cycle to the \ac{TSCH} schedule. Therefore, decreasing $N_{ch}$ reduces the required routers but compromises the battery-less device's reliability. Lack of synchronization leads to interference between battery-less devices within range. Therefore, \ac{CCA} and channel hopping, if the current channel is occupied, are essential for reliability, as a time back-off is undesirable due to energy constraints. As a result, the choice of $N_{ch}$ is a trade-off between battery-less device reliability and router performance. A practical alternative involves multi-interface routers capable of listening on all channels simultaneously. This requires only one router if the battery-less device's reliability may be slightly reduced, as the router will be incapable of receiving only when a timeslot is scheduled in the \ac{TSCH} schedule.

\subsection{Use Cases}
This approach alleviates the battery-less node from performing the energy-intensive joining procedure, while simultaneously removing the synchronization requirement. Hence, it is much more likely to successfully integrate a battery-less node in the network. Both storage and energy harvesting requirements are significantly reduced, as the energy to transmit a data frame is much lower compared to the joining procedure and periodical synchronization. In fact, as shown in Table \ref{tab:comparison}, a minimal capacitance of \SI{100}{\micro\farad} would be sufficient to transmit a data frame and the harvesting power can be as low as \SI{50}{\micro\watt} because it is only limited by the self-discharge rate of the supercapacitor \cite{VanLeemput2023}. In addition, compared to the ad-hoc joining approach, the update rate and latency can be improved to the length of a single timeslot $T_{slot}$. However, due to the performance reduction and increased energy consumption of routers, this approach is a poor choice for networks with high-performance requirements or battery-powered routers. Nonetheless, it might be the only strategy to successfully integrate extremely low-power battery-less devices in \acp{IWSN}.

\section{Performance Analysis}
\label{sec:evaluation}
\begin{figure}
    \centering
    \includegraphics[width=\linewidth]{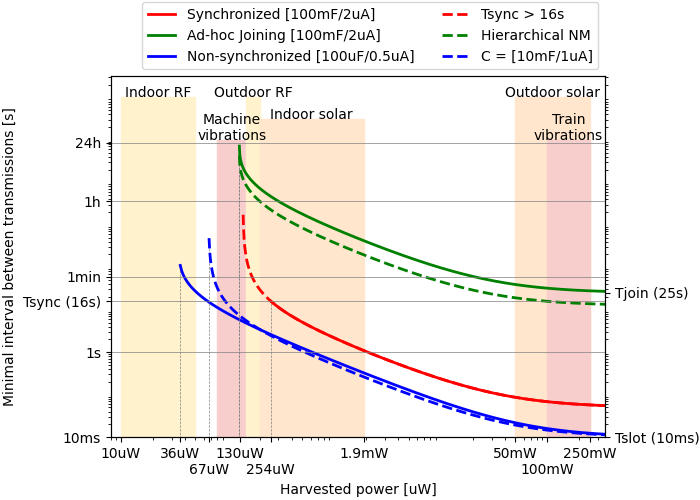}
    \caption{Analysis of the minimal interval between transmissions (lower is better) for different types of energy sources for each proposed strategy. For each strategy, the impact of one optimization is shown in dotted lines: increased synchronization interval, hierarchical network management, and increased capacitance. For the synchronized approach, train vibrations and solar panels (indoor and outdoor) are suitable energy sources, whereas devices employing the ad-hoc joining or non-synchronized approaches can also be powered using machine vibrations and outdoor \ac{RF}, depending on the desired transmission interval.}
    \label{fig:evaluation}
\end{figure}

We evaluated the proposed strategies by analyzing the network performance, energy sources, and storage capacity using the mathematical feasibility model for wireless \ac{IoT} use cases presented in Section IV of \cite{VanLeemput2023} applied to the nRF52840 platform. The model assumes a typical efficiency of \SI{80}{\percent} for both a DC-DC converter and power management unit. For more details on the model implementation and choice of parameters, we refer to Section IV and VII in \cite{VanLeemput2023} and the associated \href{https://github.com/imec-idlab/EH-feasibility}{open-source code}. Fig. \ref{fig:evaluation} depicts the achievable packet transmission interval for each approach using typical energy harvesting sources (\ac{RF}, vibration, and solar) \cite{VanLeemput2023}. (i) The synchronized approach requires a \SI{100}{\milli\farad} supercapacitor (\SI{2}{\micro\ampere} leakage) for successful joining. For a typical synchronization interval of \SI{16}{\second} (see Fig. \ref{fig:evaluation}), the lower bound of the slotframe duration is \SI{254}{\micro\watt}, corresponding to the power available from an indoor solar panel. Lower harvesting power is possible with an increased synchronization interval (red dotted line in the figure), which requires better clock accuracy and introduces additional latency.

(ii) For the ad-hoc joining strategy, the supercapacitor requirements are lower since the need for synchronization is removed. For the considered nRF platform, \SI{130}{\micro\watt} is sufficient, which is obtainable already from machine vibrations. However, this comes at the cost of an increased transmission interval (up to \SI{24}{\hour} for the lower power ranges). For more energetic energy sources, the minimal achievable latency is determined by the joining time, typically around \SI{25}{\second} \cite{Das2017, Chew2021}. Fig. \ref{fig:evaluation} also shows the impact of one discussed optimization (hierarchical network management, green dotted line in the figure), resulting in a \SI{50}{\percent} reduction in joining time \cite{Das2017}, which translates into a similar latency reduction.

(iii) The non-synchronized strategy requires only a \SI{100}{\micro\farad} capacitor (\SI{0.5}{\micro\ampere} leakage), which can already charge from \SI{36}{\micro\watt} with an acceptable transmission interval and latency of a few minutes. As a result, this approach can be used even for low-energetic energy harvesters such as indoor \ac{RF} harvesting. Due to the nature of the non-synchronized approach, the minimal achievable latency is bounded by a single transmission and acknowledgment (\SI{10}{\milli\second}). The use of a higher capacitance (\SI{10}{\milli\farad}, blue dotted line in the figure) will result in higher leakage (\SI{1}{\micro\ampere}), and will thus increase the required power and/or negatively impact the transmission interval. However, a larger capacitor also means spare capacitance and lower voltages, corresponding with increased responsiveness (see Fig. \ref{fig:intermittency}). Therefore, for energy harvesters with high harvesting power (such as solar panels), lower transmission intervals are achievable as leakage becomes negligible. As such, the choice of capacitor not only depends on the expected energy consumption but also on the expected harvesting power.

\setlength{\textfloatsep}{10pt}

\section{Conclusions}
\label{sec:conclusion}
The incorporation of battery-less devices into \acp{IWSN} offers a viable solution for extending network coverage to remote areas within industrial facilities, where the deployment of power outlets or batteries is often infeasible due to safety and maintenance considerations. However, the intermittent behavior of battery-less devices introduces two primary challenges when integrating them into \acp{IWSN}, stemming from the constrained storage capacity of supercapacitors and the unpredictability of energy harvesting sources. The first challenge, resulting from limited energy storage, poses difficulties in executing energy-intensive network joining procedures. The second challenge, arising from limited or unpredictable energy harvesting, can restrict the feasibility of frequent synchronization, a critical aspect of sustaining network connectivity. To address these challenges, this article has presented three strategies for the integration of battery-less devices into \acp{IWSN} by optimizing or eliminating the network joining and synchronization phases.

The first strategy involves a synchronized communication approach that retains both phases, maximizing \ac{IWSN} network performance without impacting router characteristics. Battery-less devices join the network during the bootstrap phase, alleviating the network's overall load. However, periodic synchronization requires consistent energy, which is often unattainable for most energy harvesters, even with adaptive synchronization and \ac{KA} frames designed to optimize synchronization. This approach is mainly suitable for energy harvesters such as indoor solar panels. The second strategy adopts an ad-hoc joining approach, which is more suitable when sustaining synchronization becomes challenging due to unreliable or limited energy harvesting. In this scenario, battery-less devices join the network for each data update, synchronizing only once. Although this approach accommodates variable energy harvesting, it comes at the cost of reduced network performance for both the battery-less device and the network. However, energy sources such as outdoor RF harvesting also become feasible. The third strategy is designed for battery-less devices equipped with extremely low-power energy harvesters such as indoor RF and machine vibrations, employing a non-synchronized communication approach that eliminates both joining and synchronization phases. While this approach imposes more lenient requirements on the battery-less device, it further diminishes network performance.

\section*{Acknowledgment}
Part of this research was funded by the Flemish FWO SBO S001521N IoBaLeT (Sustainable Internet of Batteryless Things) project.

\bibliographystyle{IEEEtran}
\bibliography{IEEEabrv,./BibTex}

\vskip -2\baselineskip plus -1fil
\begin{IEEEbiographynophoto}{Dries Van Leemput}
received his B.Sc. (2017) and M.Sc. (2018) in Electronics and ICT Engineering Technology, and M.Sc. in Electrical Engineering (2020) from Ghent University, Belgium. He began his Ph.D. studies in 2020 at Ghent University and imec, where he joined the Internet Technology and Data Science Lab (IDLab) research group. His research interests mainly revolve around multi-hop wireless sensor networks in industrial environments, Internet of Things (IoT), energy harvesting, network and energy modeling, and MAC protocol design for critical wireless systems.
\end{IEEEbiographynophoto}

\vskip -2\baselineskip plus -1fil
\begin{IEEEbiographynophoto}{Jeroen Hoebeke}
received the Masters degree in Engineering Computer Science from Ghent University in 2002. In 2007, he obtained a Ph.D. in Engineering Computer Science with his research on adaptive ad hoc routing and Virtual Private Ad Hoc Networks. Current, he is an associate professor in the Internet Technology and Data Science Lab of Ghent University and imec. He is conducting and coordinating research on wireless (IoT) connectivity, embedded communication stacks, deterministic wireless communication and wireless network management. He is author or co-author of more than 200 publications in international journals or conference proceedings.
\end{IEEEbiographynophoto}

\vskip -2\baselineskip plus -1fil
\begin{IEEEbiographynophoto}{Eli De Poorter}
is currently a Professor with the IDLab Research Group, Ghent University, and imec. His team performs research on wireless communication technologies such as (indoor) localization solutions, wireless IoT solutions, and machine learning for wireless systems. He performs both fundamental and applied research. For his fundamental research, he is currently the coordinator of several research projects (SBO, FWO, and GOA). He has over 200 publications in international journals or in the proceedings of international conferences. For his applied research, he collaborates with industry partners to transfer research results to industrial applications, and to solve challenging industrial research problems. 
\end{IEEEbiographynophoto}
\end{document}